\documentclass[12pt]{article}

\usepackage{amsmath,amssymb,amsthm}

\numberwithin{equation}{section}

\newcommand{\real}{\mathbb{R}}
\newcommand{\cS}{\mathcal{S}} 
\newcommand{\vr}{\boldsymbol{r}}
\newcommand{\vj}{\boldsymbol{j}}
\newcommand{\vv}{\boldsymbol{v}}
\newcommand{\vomega}{\boldsymbol{\omega}}
\newcommand{\Phiv}{\varPhi}
\newcommand{\Psiv}{\varPsi}

\begin{document}

\title{\bf Stationary Flows of \\
the Parabolic Potential Barrier \\
in Two Dimensions}

\author{Toshiki Shimbori$^*$ and Tsunehiro Kobayashi$^\dag$ \\
{\footnotesize\it $^*$Institute of Physics, University of Tsukuba}\\
{\footnotesize\it Ibaraki 305-8571, Japan}\\
{\footnotesize\it $^\dag$Department of General Education 
for the Hearing Impaired,}
{\footnotesize\it Tsukuba College of Technology}\\
{\footnotesize\it Ibaraki 305-0005, Japan}}

\date{}

\maketitle

\begin{abstract}
 In the two-dimensional isotropic parabolic potential barrier 
 $V(x, y)=V_0 -m\gamma^2 (x^2+y^2)/2$, 
 though it is a model of an unstable system in quantum mechanics, 
 we can obtain the stationary states 
 corresponding to the real energy eigenvalue $V_0$. 
 Further, they are infinitely degenerate. 
 For the first few eigenstates, 
 we will find the stationary flows round a right angle 
 that are expressed by the complex velocity potentials 
 $W=\pm\gamma z^2/2$. 
\end{abstract}

\thispagestyle{empty}

\setcounter{page}{0}

\pagebreak

 \section{Introduction} \label{sect.4.1.0}
 It is well known that 
 the two-dimensional harmonic oscillator 
 is equivalent to the dynamical system consisting of 
 the two independent 
 one-dimensional harmonic oscillators---the energy eigenvalues 
 of the two-dimensional oscillator are given 
 by the sum of the energy eigenvalues of the one-dimensional oscillators 
 and the eigenstates of the system are given 
 by the product of the eigenstates of the one-dimensional oscillators. 
 When degenerate eigenstates of the two-dimensional oscillator 
 are superposed with a suitable weights, 
 the new states will be the eigenstates of orbital angular momentum. 
 These results were studied a long time ago by Dirac~\cite{dirac0}. 
 
 In the present paper we will investigate 
 the two-dimensional parabolic potential barrier, 
 which is a model of an unstable system in quantum mechanics, 
 on the same lines as the two-dimensional harmonic oscillator. 
 This model is equivalent to the dynamical system consisting of 
 the two independent one-dimensional parabolic potential barriers. 
 The one-dimensional potential barrier 
 was studied by the authors~\cite{sk,s2}. 
 It is shown that 
 the energy eigenvalues are 
 complex numbers and the corresponding eigenfunctions 
 are expressible in terms of the generalized functions 
 of a Gel'fand triplet. 
 In two dimensions 
 the exact solutions of the eigenvalue problem of this model 
 separate into four types. 
 We will take the types of two of the four types in \S~\ref{sect.4.3.1}, 
 for which the solutions are 
 expressed by generalized eigenfunctions 
 belonging to complex energy eigenvalues 
 and represented by 
 diverging and converging flows. 
 In these two types the solutions will also be 
 the eigenstates of orbital angular momentum. 
 We will study the other two of the four types in \S~\ref{sect.4.3.2}. 
 In these two types all the solutions are infinitely degenerate and 
 involve the special solutions 
 with real energy eigenvalue. 
 Such special solutions are represented by 
 stationary corner flows. 
 From the hydrodynamical point of view 
 they can be described by 
 some kind of quantum ``velocity'' and 
 complex velocity potential 
 discussed in a previous paper~\cite{sk3}. 
 It has been pointed out to us by R.~Jackiw that 
 this velocity was originally introduced half a century ago 
 by Madelung~\cite{madelung}. 
 Such a velocity is still useful 
 in present-day high-energy physics~\cite{bj}. 
 We shall see that, for the first few solutions, 
 the velocities reviewed in \S~\ref{sect.4.2.0} are solenoidal, 
 so the corresponding complex velocity potentials must exist. 
 These complex velocity potentials for 
 the two-dimensional parabolic potential barrier 
 describe the flows round a right angle. 
 
 \section{Complex velocity potentials in quantum mechanics} 
 \label{sect.4.2.0}
 In the present section we shall summarize 
 the features of the velocity in quantum mechanics. 
 
 Define the velocity of a state $\psi(t,\vr)$ 
 in non-relativistic quantum mechanics by 
 \begin{equation}
  \vv\equiv\frac{\vj(t,\vr)}{\left| \psi(t,\vr)\right|^2}, 
   \label{3v.1.5}
 \end{equation}
 where $\vj(t,\vr)$ is the probability current 
 \begin{equation}
  \vj(t,\vr)\equiv\left.\Re\left[\psi(t,\vr)^*
  \left(-i\hslash\nabla\right)\psi(t,\vr)\right]\right/m, 
  \label{3.1.2}
 \end{equation}
 and $m$ is the mass of the particle. 

 If all variables are separable, 
 this velocity is in general irrotational, namely, 
 the vorticity defined by 
 \begin{equation}
  \vomega\equiv\nabla\times\vv \label{3v.1.8}
 \end{equation} 
 vanishes. 
 Then the velocity in irrotational flow 
 may be described by the gradient of the velocity potential $\Phiv$, 
 \begin{equation}
  \vv=\nabla\Phiv. \label{3v.1.9}
 \end{equation}
 
 We now proceed to study only the two-dimensional flow. 
 Let us consider the velocity \eqref{3v.1.5} 
 which is solenoidal, namely 
 \begin{equation}
  \nabla\cdot\vv\equiv 
   \frac{\partial v_x}{\partial x} 
   +\frac{\partial v_y}{\partial y}=0. \label{3v.1.11} 
 \end{equation}
 The velocity in two-dimensional flow satisfying 
 this solenoidal condition 
 may be described by the rotation of the stream function $\Psiv$, 
 \begin{equation}
  v_x=\frac{\partial\Psiv}{\partial y}, \,\,\, 
   v_y=-\frac{\partial\Psiv}{\partial x}. \label{3v.1.12}
 \end{equation}
 
 Further, in the two-dimensional irrotational flow, 
 equations \eqref{3v.1.9} and \eqref{3v.1.12} 
 can be combined into 
 Cauchy-Riemann's equations 
 between the velocity potential and the stream function. 
 We can therefore take the complex velocity potential 
 \begin{equation}
  W(z)=\Phiv(x,y)+i\Psiv(x,y), \label{3.1.13}
 \end{equation}
 which is a regular function of the complex variable $z=x+iy$. 
 For example, 
 the flow round the angle $\pi/a$ is expressed by 
 \begin{equation}
  W=A z^a, \label{3v.e.7} 
 \end{equation}
 $A$ being a number. 
 With $a=1$, 
 this expresses the uniform flow 
 of the two-dimensional plane wave in quantum mechanics. 
 There are some elementary examples of complex velocity potentials 
 which express 
 the two-dimensional flows in quantum mechanics~\cite{sk3}. 

 The above-mentioned method will be applied to 
 the stationary flows of the two-dimensional parabolic potential barrier 
 in \S~\ref{sect.4.3.2}. 

 \section{The parabolic potential barrier in two dimensions} 
 \label{sect.4.3.0} 
 The Hamiltonian of the two-dimensional isotropic 
 parabolic potential barrier is 
 \begin{equation}
  \hat{H}=-\frac{\hslash^2}{2m} 
   \left(\frac{\partial^2}{\partial x^2}
    +\frac{\partial^2}{\partial y^2}\right) 
    +V_0 -\frac{1}{2} m\gamma^2 \left(x^2+y^2\right), \label{4.2.1}
 \end{equation}
 where $V_0\in\real$ is the maximum potential energy, 
 $m>0$ is the mass and $\gamma>0$ is 
 proportional to the square root of the curvature at $(x,y)=(0,0)$. 
 
 A state is represented by a wave function $U(x,y)$ 
 satisfying the Schr\"{o}dinger equation, which now reads, 
 with $\hat{H}$ given by \eqref{4.2.1}, 
 \begin{equation}
  -\frac{\hslash^2}{2m} 
   \left(\frac{\partial^2}{\partial x^2}
    +\frac{\partial^2}{\partial y^2}\right) U(x,y) 
    +\left\{V_0 -\frac{1}{2} m\gamma^2 \left(x^2+y^2\right)\right\} U(x,y) 
    =E U(x,y). \label{4.2.2} 
 \end{equation}
 The energy eigenvalues of \eqref{4.2.2} will be 
 the sum of the energy eigenvalues of the one-dimensional 
 parabolic potential barrier in the $x$-direction and $y$-direction, 
 respectively, i.e. 
 \begin{equation}
  E_{n_x n_y}=E_{n_x} +E_{n_y} \label{4.2.3} 
 \end{equation}
 and the eigenfunctions belonging to these energy eigenvalues will be 
 the product of their corresponding eigenfunctions 
 \begin{equation}
  U_{n_x n_y}(x,y)=u_{n_x}(x) u_{n_y}(y). \label{4.2.4} 
 \end{equation}
 
 With the notation of preceding papers~\cite{sk,s2}, 
 the energy eigenvalues of the one-dimensional 
 parabolic potential barrier are 
 \begin{equation}
  E^\pm_{n_q} =\frac{1}{2}V_0 
   \mp i\left(n_q +\frac{1}{2}\right)\hslash\gamma \,\,\, 
   \left(n_q =0, 1, 2, \cdots\right) \label{4.2.5} 
 \end{equation}
 and the corresponding eigenfunctions are 
 \begin{equation}
  u^\pm_{n_q}(q)=e^{\pm i\beta^2 q^2/2} H^\pm_{n_q}(\beta q) \,\,\, 
   \left(\beta\equiv\sqrt{m\gamma/\hslash}\right), 
   \label{4.2.6}
 \end{equation}
 where $H^\pm_{n_q}(\beta q)$ are the polynomials of degree $n_q$, and 
 the numerical coefficients are discarded. 
 The eigenfunctions $u^\pm_{n_q}$ are generalized functions 
 in ${\cS(\real)}^\times$ of 
 the following Gel'fand triplet, 
 \begin{equation}
  \cS(\real)\subset L^2(\real)\subset{\cS(\real)}^\times,  
   \label{2.2.15}
 \end{equation}
 where $L^2(\real)$ is a Lebesgue space and 
 $\cS(\real)$ is a Schwartz space. 
 The work~\cite{sk} also shows that 
 the index $+$ means only outward moving particles and 
 the index $-$ means only inward moving particles. 
 
 Thus the results \eqref{4.2.3} and \eqref{4.2.4} 
 of the two-dimensional parabolic potential barrier 
 separate into four types: 
 \begin{xalignat*}{2}
  \text{{\it Type} 1. } && 
  &E^{++}_{n_x n_y}
  = E^+_{n_x}+E^+_{n_y}
  = V_0 -i(n_x +n_y +1)\hslash\gamma, \\*
  && 
  &U^{++}_{n_x n_y}(x,y)
  = u^+_{n_x}(x) u^+_{n_y}(y)
  = e^{+i\beta^2(x^2+y^2)/2} 
  H^+_{n_x}(\beta x) H^+_{n_y}(\beta y). \\[6pt]
  \text{{\it Type} 2. } && 
  &E^{+-}_{n_x n_y}
  = E^+_{n_x}+E^-_{n_y}
  = V_0 -i(n_x -n_y)\hslash\gamma, \\*
  && 
  &U^{+-}_{n_x n_y}(x,y)
  = u^+_{n_x}(x) u^-_{n_y}(y)
  = e^{+i\beta^2(x^2-y^2)/2} 
  H^+_{n_x}(\beta x) H^-_{n_y}(\beta y). \\[6pt]
  \text{{\it Type} 3. } && 
  &E^{-+}_{n_x n_y}
  = E^-_{n_x}+E^+_{n_y}
  = V_0 +i(n_x -n_y)\hslash\gamma, \\*
  && 
  &U^{-+}_{n_x n_y}(x,y)
  = u^-_{n_x}(x) u^+_{n_y}(y)
  = e^{-i\beta^2(x^2-y^2)/2} 
  H^-_{n_x}(\beta x) H^+_{n_y}(\beta y). \\[6pt]
  \text{{\it Type} 4. } && 
  &E^{--}_{n_x n_y}
  = E^-_{n_x}+E^-_{n_y}
  = V_0 +i(n_x +n_y +1)\hslash\gamma, \\*
  &&
  &U^{--}_{n_x n_y}(x,y)
  = u^-_{n_x}(x) u^-_{n_y}(y)
  = e^{-i\beta^2(x^2+y^2)/2} 
  H^-_{n_x}(\beta x) H^-_{n_y}(\beta y). 
 \end{xalignat*}
 These eigenfunctions $U^{+ +}_{n_x n_y}$, $U^{+ -}_{n_x n_y}$, 
 $U^{- +}_{n_x n_y}$, $U^{- -}_{n_x n_y}$ are also generalized functions 
 in ${\cS(\real^2)}^\times$ of the Gel'fand triplet 
 \begin{equation}
  \cS(\real^2)\subset L^2(\real^2)\subset{\cS(\real^2)}^\times 
   \label{4.g.1}
 \end{equation}
 instead of \eqref{2.2.15}. 
 Note that the eigenfunctions of types 4 and 3 
 are conjugate complex functions of types 1 and 2, respectively, 
 i.e. 
 $$U^{\pm\pm}_{n_x n_y}(x,y)^*=U^{\mp\mp}_{n_x n_y}(x,y)$$ 
 and 
 $$U^{\pm\mp}_{n_x n_y}(x,y)^*=U^{\mp\pm}_{n_x n_y}(x,y). $$ 

  \subsection{Diverging and converging flows}
  \label{sect.4.3.1}
\begin{figure}
   \begin{center}
    \begin{picture}(200,200)
     \thicklines
     \put(0,100){\vector(1,0){200}}
     \put(100,0){\vector(0,1){200}}
     \put(90,88){$0$}
     \put(205,98){$x$}
     \put(98,205){$y$}
     
     \put(105,105){\vector(1,1){90}}
     \put(95,105){\vector(-1,1){90}}
     \put(85,85){\vector(-1,-1){80}}
     \put(105,95){\vector(1,-1){90}}
    \end{picture}
   \end{center}
   \caption[]{Diverging flows.}
   \label{fig:4.1}
  \end{figure}
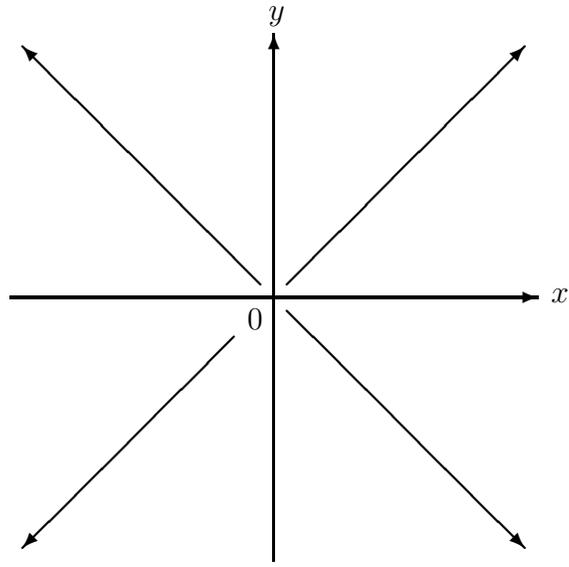

\begin{figure}
   \begin{center}
    \begin{picture}(200,200)
     \thicklines
     \put(0,100){\vector(1,0){200}}
     \put(100,0){\vector(0,1){200}}
     \put(90,88){$0$}
     \put(205,98){$x$}
     \put(98,205){$y$}
     
     \put(195,195){\vector(-1,-1){90}}
     \put(5,195){\vector(1,-1){90}}
     \put(5,5){\vector(1,1){80}}
     \put(195,5){\vector(-1,1){90}}
    \end{picture}
   \end{center}
   \caption[]{Converging flows.}
   \label{fig:4.2}
  \end{figure}
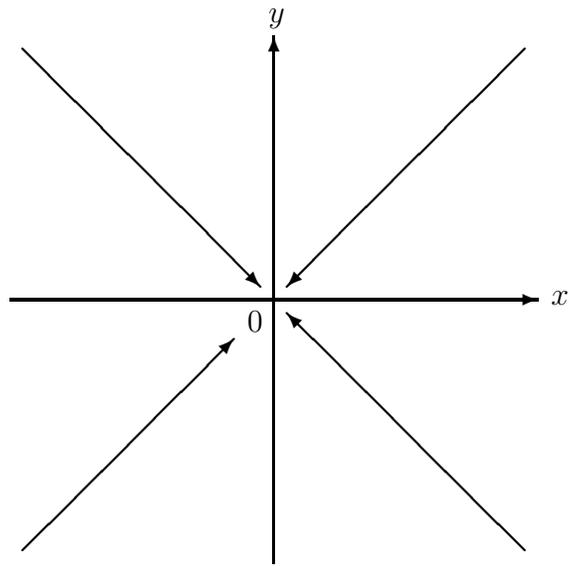

  Let us consider first the types 1 and 4. 
  In this case the energy eigenvalues $E^{\pm\pm}_{n_x n_y}$ are 
  always complex numbers 
  and the time factors corresponding to them are 
  $$e^{-i E^{\pm\pm}_{n_x n_y} t/\hslash}
  =e^{-iV_0t/\hslash}e^{\mp(n_x+n_y+1)\gamma t}. $$ 
  Thus the solutions of type 1 are well defined when $t>0$, 
  and those of type 4 are well defined when $t<0$, 
  according to the time boundary condition 
  that time factors of an unstable system 
  are square integrable~\cite{sk}. 
  Also, 
  $U^{+ +}_{n_x n_y}(x,y)$ represent particles 
  moving outward from the center as in fig.~\ref{fig:4.1},  
  and 
  $U^{- -}_{n_x n_y}(x,y)$ represent particles 
  moving inward to the center as in fig.~\ref{fig:4.2}.
  Thus we shall call these types {\it diverging} 
  and {\it converging flows}, 
  respectively. 
  Note that a time reversal occurs 
  resulting in the interchange of the diverging 
  and converging flows. 
  
  For $n_x=n_y=0$, 
  we get the energy eigenvalue 
  \begin{equation}
   E^{\pm\pm}_{0 0}=V_0\mp i\hslash\gamma \label{4.2.7} 
  \end{equation}
  and only one eigenfunction
  \begin{equation}
   U^{\pm\pm}_{0 0}(x,y)=e^{\pm i\beta^2(x^2+y^2)/2}, \label{4.2.8}
  \end{equation}
  respectively. 
  For $n_x+n_y=1$, namely $n_x=1, n_y=0$ and $n_x=0, n_y=1$, 
  we get 
  \begin{equation}
   E^{\pm\pm}_{1 0}=E^{\pm\pm}_{0 1}=V_0\mp 2i\hslash\gamma \label{4.2.9} 
  \end{equation}
  and two eigenfunctions 
  \begin{equation}
   \left.
    \begin{aligned}
     U^{\pm\pm}_{1 0}(x,y)&=2\beta x e^{\pm i\beta^2(x^2+y^2)/2}, \\
     U^{\pm\pm}_{0 1}(x,y)&=2\beta y e^{\pm i\beta^2(x^2+y^2)/2}. 
    \end{aligned} \right\} \label{4.2.10}
  \end{equation}
  There is twofold degenerate state of types 1 and 4 
  with $n_x+n_y=1$. 
  For $n_x+n_y=2$, namely $n_x=2, n_y=0$; $n_x=1, n_y=1$; $n_x=0, n_y=2$, 
  we get 
  \begin{equation}
   E^{\pm\pm}_{2 0}=E^{\pm\pm}_{1 1}=E^{\pm\pm}_{0 2}
    =V_0\mp 3i\hslash\gamma \label{4.2.11} 
  \end{equation}
  and three eigenfunctions 
  \begin{equation}
   \left.
    \begin{aligned}
     U^{\pm\pm}_{2 0}(x,y)
     &=(4\beta^2 x^2\mp 2i) e^{\pm i\beta^2(x^2+y^2)/2}, \\
     U^{\pm\pm}_{1 1}(x,y)
     &=4\beta^2 xy e^{\pm i\beta^2(x^2+y^2)/2}, \\
     U^{\pm\pm}_{0 2}(x,y)
     &=(4\beta^2 y^2\mp 2i) e^{\pm i\beta^2(x^2+y^2)/2}.
    \end{aligned} \right\} \label{4.2.12}
  \end{equation}
  There is threefold degenerate state of types 1 and 4 
  with $n_x+n_y=2$. 
  Generally, there is $(n+1)$-fold degenerate state of types 1 and 4 
  with $n_x+n_y=n$. This result is just the same degree of degeneracy as 
  the two-dimensional harmonic oscillator. 
  
  For the further discussion of the state of types 1 and 4, 
  we now pass from the Cartesian coordinates $x, y$ 
  to the two-dimensional polar coordinates $r, \varphi$ 
  by means of the equations 
  \begin{equation}
   \left.
    \begin{aligned}
     x&=r\cos\varphi, \\
     y&=r\sin\varphi. 
    \end{aligned} \right\} \label{4.2.13}
  \end{equation} 
  If in the new coordinates we superpose above-mentioned eigenstates 
  with suitable weights, 
  the result will be the eigenstates of 
  orbital angular momentum $\hat{L}$ defined by 
  \begin{equation}
   \hat{L}=-i\hslash
    \left(x\frac{\partial}{\partial y}-y\frac{\partial}{\partial x}\right)
    =-i\hslash \frac{\partial}{\partial\varphi}. \label{4.2.14} 
  \end{equation}
  For $n_x=n_y=0$, 
  the eigenfunction \eqref{4.2.8} will be 
  \begin{equation}
   U^{\pm\pm}_{00}(r,\varphi)=e^{\pm i\beta^2 r^2/2}. \label{4.2.15} 
  \end{equation}
  Thus $U^{\pm\pm}_{00}(r,\varphi)$ is independent of $\varphi$ and 
  has zero orbital angular momentum. 
  For $n_x+n_y=1$, 
  a linear combination of the eigenfunctions \eqref{4.2.10} give 
  \begin{equation}
   \left.
    \begin{aligned}
     U^{\pm\pm}_{10}(r,\varphi)+iU^{\pm\pm}_{01}(r,\varphi) 
     &=2\beta r e^{\pm i\beta^2 r^2/2} e^{i\varphi}, \\
     U^{\pm\pm}_{10}(r,\varphi)-iU^{\pm\pm}_{01}(r,\varphi)
     &=2\beta r e^{\pm i\beta^2 r^2/2} e^{-i\varphi}. 
    \end{aligned} \right\} \label{4.2.16}
  \end{equation}
  These states are eigenstates of $\hat{L}$ 
  with eigenvalues $\hslash$ and $-\hslash$, respectively. 
  For $n_x+n_y=2$, 
  \eqref{4.2.12} give 
  \begin{equation}
   \left.
    \begin{aligned}
     U^{\pm\pm}_{20}(r,\varphi)
     +2iU^{\pm\pm}_{11}(r,\varphi)-U^{\pm\pm}_{02}(r,\varphi) 
     &=4\beta^2 r^2 e^{\pm i\beta^2 r^2/2} e^{2i\varphi}, \\
     U^{\pm\pm}_{20}(r,\varphi)
     +U^{\pm\pm}_{02}(r,\varphi)
     &=4(\beta^2 r^2 \mp i) e^{\pm i\beta^2 r^2/2}, \\
     U^{\pm\pm}_{20}(r,\varphi)
     -2iU^{\pm\pm}_{11}(r,\varphi)-U^{\pm\pm}_{02}(r,\varphi) 
     &=4\beta^2 r^2 e^{\pm i\beta^2 r^2/2} e^{-2i\varphi}. 
    \end{aligned} \right\} \label{4.2.17}
  \end{equation}
  These states are also eigenstates of $\hat{L}$ 
  with eigenvalues $2\hslash$, $0$, and $-2\hslash$. 
  There is a similar procedure for large values of $n_x+n_y$. 
  
  \subsection{Corner flows} \label{sect.4.3.2}
\begin{figure}[pb]
   \begin{center}
    \begin{picture}(200,200)
     \thicklines
     \put(0,100){\vector(1,0){200}}
     \put(100,0){\vector(0,1){200}}
     \put(90,88){$0$}
     \put(205,98){$x$}
     \put(98,205){$y$}
     
     \qbezier(105,200)(105,105)(200,105)
     \qbezier(95,200)(95,105)(0,105)
     \qbezier(95,0)(95,95)(0,95)
     \qbezier(105,0)(105,95)(200,95)
     
     \put(200,105){\vector(1,0){1}}
     \put(0,105){\vector(-1,0){1}}
     \put(0,95){\vector(-1,0){1}}
     \put(200,95){\vector(1,0){1}}
    \end{picture}
   \end{center}
   \caption[]{Corner flows moving from the $y$-direction 
   to the $x$-direction.}
   \label{fig:4.3}
  \end{figure}
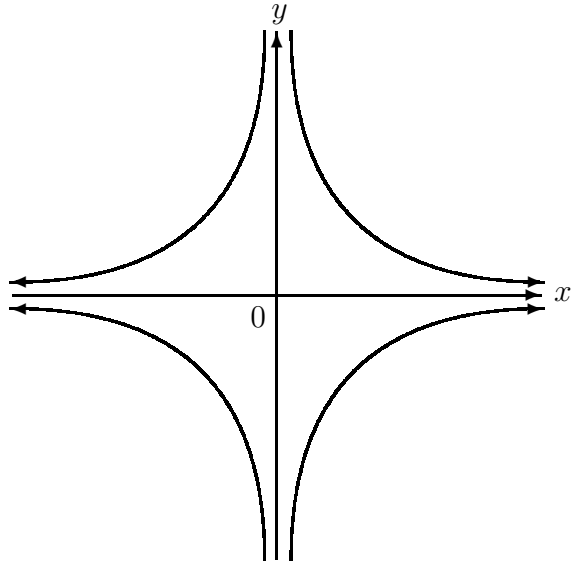

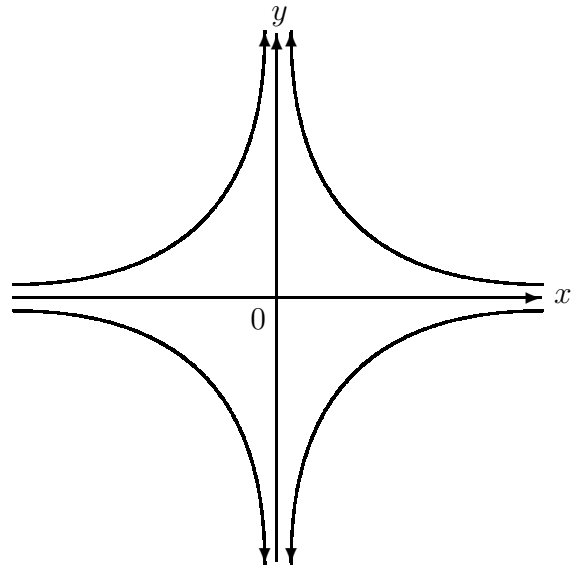
\begin{figure}
   \begin{center}
    \begin{picture}(200,200)
     \thicklines
     \put(0,100){\vector(1,0){200}}
     \put(100,0){\vector(0,1){200}}
     \put(90,88){$0$}
     \put(205,98){$x$}
     \put(98,205){$y$}
     
     \qbezier(105,200)(105,105)(200,105)
     \qbezier(95,200)(95,105)(0,105)
     \qbezier(95,0)(95,95)(0,95)
     \qbezier(105,0)(105,95)(200,95)
     
     \put(105.5,200){\vector(0,1){1}}
     \put(95.5,200){\vector(0,1){1}}
     \put(95.5,0){\vector(0,-1){1}}
     \put(105.5,0){\vector(0,-1){1}}
    \end{picture}
   \end{center}
   \caption[]{Corner flows moving from the $x$-direction 
   to the $y$-direction.}
   \label{fig:4.4}
  \end{figure}

  Let us now study the types 2 and 3. 
  In this case the energy eigenvalues $E^{\pm\mp}_{n_x n_y}$ are 
  also in general complex numbers, 
  but with the striking difference that 
  all the eigenstates belonging to each energy eigenvalue are 
  infinitely degenerate. 
  The corresponding time factors are 
  $$e^{-i E^{\pm\mp}_{n_x n_y} t/\hslash}
  =e^{-iV_0t/\hslash}e^{\mp(n_x-n_y)\gamma t}. $$ 
  Thus the solutions of type 2 are well defined when $t>0$, 
  and those of type 3 are well defined when $t<0$, 
  for the case of $n_x>n_y$, and vice versa. 
  Also, 
  $U^{+ -}_{n_x n_y}(x,y)$ represent particles 
  which, coming from the $y$-direction, 
  round the center and, 
  go off to the $x$-direction as in fig.~\ref{fig:4.3}, 
  and 
  $U^{- +}_{n_x n_y}(x,y)$ represent particles  
  which, coming from the $x$-direction, 
  round the center and, 
  go off to the $y$-direction as in fig.~\ref{fig:4.4}. 
  Note that a time reversal occurs 
  resulting in the interchange of these corner flows. 
  
    \paragraph{Stationary flows \\}
    The above time factors now show that 
    for the case of $n_x=n_y$, there are {\it stationary flows}. 
    For $n_x=n_y=n=0, 1, 2, \cdots ,$ 
    the energy eigenvalues associated with stationary flows are 
    the same real number: 
    \begin{equation}
     E^{\pm\mp}_{n n}=V_0. \label{4.2.18} 
    \end{equation}
    The first few infinitely degenerate eigenfunctions 
    belonging to this energy eigenvalue \eqref{4.2.18} are 
    \begin{equation}
     \left.
      \begin{aligned}
       U^{\pm\mp}_{0 0}(x,y)&=e^{\pm i\beta^2(x^2-y^2)/2}, \\
       U^{\pm\mp}_{1 1}(x,y)&=4\beta^2 xy e^{\pm i\beta^2(x^2-y^2)/2}, \\
       U^{\pm\mp}_{2 2}(x,y)&=
       4\left[4\beta^4 x^2 y^2 +1\pm 2i\beta^2(x^2-y^2)\right] 
       e^{\pm i\beta^2(x^2-y^2)/2}, \\
       &\dots .
      \end{aligned} \right\} \label{4.2.19}
    \end{equation}
    
    For the further study of the stationary flows 
    of the two-dimensional parabolic potential barrier 
    with the Hamiltonian \eqref{4.2.1}, 
    it is convenient to make a transformation to 
    the rectangular hyperbolic coordinates $u, v$, given by 
    \begin{equation}
     \left.
      \begin{aligned}
       u&=x^2-y^2, \\
       v&=2xy. 
      \end{aligned} \right\} \label{4.2.20}
    \end{equation}
    The eigenfunctions \eqref{4.2.19} will become 
    in the new representation 
    \begin{equation}
     \left.
      \begin{aligned}
       U^{\pm\mp}_{0 0}(u,v)&=e^{\pm i\beta^2 u/2}, \\
       U^{\pm\mp}_{1 1}(u,v)&=2\beta^2 v e^{\pm i\beta^2 u/2}, \\
       U^{\pm\mp}_{2 2}(u,v)&=
       4\left(\beta^4 v^2 +1\pm 2i\beta^2 u\right) 
       e^{\pm i\beta^2 u/2}, \\
       &\dots .
      \end{aligned} \right\} \label{4.2.21}
    \end{equation}
    The factors $e^{\pm i\beta^2 u/2}$ occurred in \eqref{4.2.21} 
    describe plane waves in the $uv$-plane, i.e. 
    the motion of the wave $e^{i\beta^2 u/2}$ 
    is in the direction specified by fig.~\ref{fig:4.3} and 
    that of the wave $e^{-i\beta^2 u/2}$ 
    is in the direction specified by fig.~\ref{fig:4.4}. 
    These eigenfunctions substituted in \eqref{3.1.2} 
    give the rectangular hyperbolic coordinates 
    $j^{\pm\mp}_{n n u}$, $j^{\pm\mp}_{n n v}$ of 
    $\vj^{\pm\mp}_{n n}$, which are the probability currents 
    of the states $U^{\pm\mp}_{n n}$. 
    They give 
    \begin{equation*}
      \begin{aligned}
       j^{\pm\mp}_{0 0 u}(u,v)&=\pm\gamma h_u/2,\,\,\, 
       j^{\pm\mp}_{0 0 v}(u,v)=0, \\
       j^{\pm\mp}_{1 1 u}(u,v)
       &=\pm 2\gamma\beta^4 v^2 h_u,\,\,\,  
       j^{\pm\mp}_{1 1 v}(u,v)=0, \\
       j^{\pm\mp}_{2 2 u}(u,v)
       &=\pm 8\gamma 
       \left\{(\beta^4 v^2 +5)(\beta^4 v^2 +1)+4\beta^4 u^2\right\} 
       h_u, \\ 
       j^{\pm\mp}_{2 2 v}(u,v)
       &=\mp 64\gamma\beta^4 uv h_v, \\ 
       &\dots ,
      \end{aligned} 
    \end{equation*}
    where the scale factors $h_u=h_v=2\sqrt[4]{u^2+v^2}$. 
    Thus $\vj^{\pm\mp}_{n n}$ can never depend on the time $t$. 
    We see from this result the suitability of the term 
    ``stationary flows''. 

    \paragraph{The flows round a right angle \\}
    The above probability currents now show that 
    for the case of $n=0\text{ and }1$, 
    there are stationary flows which move along 
    the hyperbolas (each line with $v$ constant). 
    To get an understanding of the physical features of this flows 
    it is better to work with the velocity defined by \eqref{3v.1.5}. 
    For $n=0\text{ and }1$, the velocities give the same result 
    $$v^{\pm\mp}_u=\pm\frac{1}{2}\gamma h_u,\,\,\, 
    v^{\pm\mp}_v=0. $$
    Taking the rotation of them, 
    the vorticity \eqref{3v.1.8} becomes 
    $$\omega^{\pm\mp}\equiv h_u h_v 
    \left[\frac{\partial}{\partial u}
    \left(\frac{v^{\pm\mp}_v}{h_v}\right)
    -\frac{\partial}{\partial v}
    \left(\frac{v^{\pm\mp}_u}{h_u}\right)\right]=0. $$   
    These equations will hold generally in quantum mechanics~\cite{sk3}, 
    and therefore the velocity potentials defined by \eqref{3v.1.9} 
    must exist. 
    If we transform to rectangular hyperbolic coordinates $u$, $v$, 
    equations \eqref{3v.1.9} become 
    \begin{equation}
     v_u=h_u \frac{\partial\Phiv}{\partial u},\,\,\, 
      v_v=h_v \frac{\partial\Phiv}{\partial v}, \label{4.p.9} 
    \end{equation}
    and the velocity potentials for $n=0\text{ and }1$ are thus 
    \begin{equation}
     \Phiv^{\pm\mp}=\pm\frac{1}{2}\gamma u. \label{4.3.3} 
    \end{equation}
    Note that they are proportional to the phase factors of \eqref{4.2.21}. 
    Further, the divergence \eqref{3v.1.11} gives 
    $$\nabla\cdot\vv^{\pm\mp}\equiv h_u h_v 
    \left[\frac{\partial}{\partial u}
    \left(\frac{v^{\pm\mp}_u}{h_v}\right)
    +\frac{\partial}{\partial v}
    \left(\frac{v^{\pm\mp}_v}{h_u}\right)\right]=0. $$
    Thus the velocities are  solenoidal for $n=0\text{ and }1$, 
    so that we can obtain the stream functions 
    defined by \eqref{3v.1.12}. 
    The equations \eqref{3v.1.12} are also expressed, 
    as in equations \eqref{4.p.9}, 
    \begin{equation}
     v_u=h_v \frac{\partial\Psiv}{\partial v},\,\,\, 
      v_v=-h_u \frac{\partial\Psiv}{\partial u}, \label{4.p.12} 
    \end{equation}
    and the stream functions for $n=0\text{ and }1$ are thus 
    \begin{equation}
     \Psiv^{\pm\mp}=\pm\frac{1}{2}\gamma v. \label{4.3.5} 
    \end{equation}
    For the states represented by 
    the first and second of equations \eqref{4.2.19} or \eqref{4.2.21}, 
    the complex velocity potential \eqref{3.1.13} gives, 
    from \eqref{4.3.3} and \eqref{4.3.5} 
    \begin{align}
     W^{\pm\mp}&=\pm\frac{1}{2}\gamma u \pm\frac{i}{2}\gamma v \notag \\
     &=\pm\frac{1}{2}\gamma z^2, \label{4.3.6} 
    \end{align}
    since $z^2=u+iv$. 
    Equations \eqref{4.3.6} are of the form \eqref{3v.e.7} with $a=2$, 
    and they show that, {\it for $n=0\text{ and }1$, 
    the complex velocity potentials 
    of the two-dimensional parabolic potential barrier express 
    the flows round a right angle}. 
    One could work out in terms of Cartesian coordinates 
    and one would be led to the same conclusion. 
    
 \section{Discussion} \label{sect.4.4.0} 
 We have obtained the exact solutions 
 of the two-dimensional parabolic potential barrier. 
 One class of the solutions is diverging and converging flows 
 of \S~\ref{sect.4.3.1}. 
 These solutions are always complex energy eigenvalues and 
 generalized eigenfunctions, 
 which mean that the diverging and converging flows are 
 not stationary. 
 These generalized eigenfunctions can be obtained from 
 the eigenfunctions of the two-dimensional harmonic oscillator 
 by the analytical continuation, 
 in the same way as 
 the one-dimensional parabolic potential barrier~\cite{sk}. 
 Again, 
 they can be superposed to give 
 the eigenstates of orbital angular momentum. 
 For these solutions, however, 
 the method mentioned in \S~\ref{sect.4.2.0} was not applicable. 
 As an example we try to calculate the divergence \eqref{3v.1.11} 
 for the states \eqref{4.2.8} or \eqref{4.2.15}. 
 The result is 
 $$\nabla\cdot\vv^{\pm\pm}_{0 0}=\pm 2\gamma \neq 0. $$
 Now the $\pm$ sign shows that 
 $U^{++}_{0 0}$ is connected with the diverging flow and 
 $U^{--}_{0 0}$ is connected with the converging one.  
 Thus the velocities of diverging and converging flows 
 cannot be solenoidal. 
 Therefore the stream functions and 
 the complex velocity potentials do not exist. 
 This result still holds for large values of $n_x+n_y$. 
 
 The other class of the solutions is corner flows 
 of \S~\ref{sect.4.3.2}. 
 All the solutions are infinitely degenerate and 
 involve the stationary flows with real energy eigenvalue. 
 It should be noted that there are no stationary flows 
 in the one-dimensional or three-dimensional 
 {\it isotropic} parabolic potential barrier. 
 For $n=0\text{ and }1$ in the stationary flows, 
 we have found the flows round a right angle that are expressed 
 by the complex velocity potentials \eqref{4.3.6}. 
 But for $n\geqslant 2$, 
 the complex velocity potentials do not exist, 
 because the imaginary parts of the polynomials $H^\pm_{n_q}(\beta q)$ 
 in \eqref{4.2.6} cause the stream lines to depart from hyperbolas. 
 One may, however, find a new flow as the result of a kind of 
 superposition of the infinitely degenerate states. 
 
 One would expect to be able to get a more direct solution 
 of the eigenvalue problem of the Hamiltonian \eqref{4.2.1} 
 by working all the time in the two-dimensional polar coordinates, 
 instead of working in the Cartesian coordinates and 
 transforming at the end to the two-dimensional polar coordinates, 
 as was done in \S~\ref{sect.4.3.1}. 
 But under suitable boundary conditions 
 in the two-dimensional polar coordinates, 
 one would obtain only diverging and converging flows of \S~\ref{sect.4.3.1}, 
 i.e. the lack of corner flows of \S~\ref{sect.4.3.2}. 
 It is also pointed out that 
 one can get the only diverging and converging flows 
 from the analytical continuation of the solutions 
 of the two-dimensional harmonic oscillator. 
 These facts mean that 
 {\it the choice of coordinate systems is quite important 
 in the eigenvalue problem of 
 the unstable system in non-relativistic quantum mechanics, 
 since coordinate systems impose a restriction on 
 the symmetry of boundary conditions}. 
 The source of the conclusion lies in the existence of 
 a very large class of solutions for the unstable system. 
 
 It is rather surprising that such a stable idea as 
 stationary flows should appear in 
 the parabolic potential barrier in this way. 
 Actually we have shown in another paper~\cite{ks5} that 
 the dynamical system composed of several of these potential barriers 
 forms the quasi-stable semiclassical system.

\end{document}